\documentclass[prl,showpacs,twocolumn,amsmath,amssymb,a4paper,floatfix]{revtex4}

\usepackage{graphicx}% Include figure files
\usepackage{dcolumn}% Align table columns on decimal point
\usepackage{bm}% bold math

\begin{document}

\title{Magnetism in Exact Exchange Density Functional Theory}

\author{S. Sharma}
\email{sangeeta.sharma@uni-graz.at}
\author{J.~K. Dewhurst}
\author{S.~A. Sagmeister }
\author{C. Ambrosch-Draxl}
\affiliation{Institut f\"ur Physik, Karl-Franzens-Universit\"at Graz, \\
Universit\"atsplatz 5, A-8010 Graz, Austria.}
\author{C. Persson}
\affiliation{Department of Materials Science and Engineering, \\
Royal Institute of Technology, SE-100 44 Stockholm, Sweden}
\author{S. Shallcross}
\affiliation{Department of Physics and Measurement Technology, \\
University of Link\"oping, SE-581 83 Link\"oping, Sweden.}
\author{L. Nordstr\"om}
\affiliation{Department of Physics, Uppsala University, \\
Box 530, 751 21 Uppsala, Sweden.}

\date{\today}

\begin{abstract}
The magnetic properties of the  intermetallic compound FeAl are investigated
using exact exchange density functional theory. This is implemented
within a state of the art all-electron full potential method. We find that 
FeAl is magnetic with a moment of 0.70 $\mu_B$, close to the LSDA result of 
0.69 $\mu_B$. A comparison with the non-magnetic density of states with 
experimental negative binding energy result shows a much better agreement 
than any previous calculations. We attribute this to the fine details
of the exchange field, in particular its asymmetry, which is captured very
well with the orbital dependent exchange potential.
\end{abstract}

                             
\maketitle

\section{Introduction}

The intermetallic compound FeAl presents an interesting system where one has both
a complex interplay between structure and magnetism, and a possible failure of
the Local Spin Density Approximation (LSDA) to adequately account for the electronic structure
\cite{pet03,smir05,bog98}. Disentangling these two strands poses a severe challenge
for theory. The controversy surrounding this material turns on the issue of whether
\emph{completely ordered} FeAl is a magnetic or non-magnetic intermetallic, and whether
this question is at all relevant to the experimental picture.

Experimentally, FeAl obeys an extended Curie-Weiss law 
$(\chi-\chi_0)^{-1} = C(T - \Theta)$ down to low temperatures, the signature of 
a paramagnet with finite (local) moments. M\"ossbauer studies \cite{bog98} show 
that it is only the antisite Fe atoms (Fe atoms on the Al sublattice) which 
contribute to this behaviour, the Fe sublattice being non-magnetic. The rather 
weak ordering in FeAl means both that a range of defects are thermodynamically 
stable, and that any crystal of FeAl will have some concentration of these defects.
Thus it appears that the question of whether completely ordered FeAl is magnetic 
or not cannot be answered by experiment.

Nevertheless, the question of the magnetic state of completely ordered FeAl has 
attracted a lot of attention \cite{pet03,mohn01,smir05,bog98,papa00,shab01}. 
This is based on the speculation that it should be non-magnetic \cite{mohn01}, 
and the fact that LSDA calculations find it to be magnetic with a moment of around 
0.69 $\mu_B$. The foundation of this speculation appears to be simply that since 
the magnetism is carried by defects, taking these defects out will leave behind 
the non-magnetic FeAl sublattice, hence non-magnetic FeAl. However, this excludes 
that an Fe antisite-Fe sublattice \emph{interaction} may be responsible for making
the Fe sublattice non-magnetic.

The first attempt to go beyond the LSDA was undertaken by Mohn \emph{et. al} who 
employed the LDA+U method. The usefulness of this approach was suggested by the 
fact that both the $e_g$ and $t_{2g}$ bands are rather flat, and hence an improved 
scheme for the
correlation is needed. Although these authors found a non-magnetic solution, a more
recent LDA+U calculation \cite{pet03} disagrees with this result and finds, again, 
FeAl to be magnetic. The latter work instead suggested that the magnetism may be 
suppressed by critical spin fluctuations, as is seen for example in the intermetallic 
compound Ni$_3$Ga \cite{agu04}. The dynamical version of LDA+U, the dynamical 
mean field theory (DMFT)\cite{pet03}, does include some spin-fluctuations and using 
this it was found that FeAl is non-magnetic.

In complete contrast to these approaches Smirnov \emph{et. al} \cite{smir05} instead 
attempted a more realistic modeling of the experimental defect structure within 
the LSDA. To achieve this they  employed the Coherent Potential Approximation (CPA), 
which allows for the treatment of both structural and magnetic disorder. With an 
Fe antisite concentration similar to that deduced experimentally they found that 
a spin disordered local moment (DLM) solution is then energetically favoured. 
Crucially, in this state the Fe sublattice is non-magnetic, with the magnetism 
carried only by the antisite atoms, exactly as seen in experiment.

Although it is clear that the complex defect structure must be taken into account 
in order to model correctly the experimental reality, one can note that it is not
clear that the magnetism of FeAl is correctly described by the DLM state. Evidence 
for this comes from the fact that Fe clusters in FeAl order with a non-collinear 
structure. Thus the Fe antisite-Fe sublattice interaction is not ferromagnetic, 
and hence the DLM state should be lower in energy. This does not, however, rule 
out the possibility that  some complex non-collinear structure is even lower in 
energy, and that in this state the Fe lattice is again magnetic.

In this situation, one might wish for a more experimentally rigorous comparison 
than simply whether FeAl is non-magnetic or not. Unfortunately, and perhaps 
surprisingly, there is a lack of experimental spectroscopic data for this system. 
To the best of our knowledge, only one experimental article on this aspect of 
FeAl is to be found in the literature \cite{shab01}.

In this work we deploy the EXact eXchange (EXX) method, implemented within
an all-electron full-potential code EXITING \cite{trog05}, to study both the 
magnetism and spectroscopic properties of FeAl.

\section{Method of calculation}
Recently we have developed EXX equations suitable for non-collinear 
magnets \cite{shar05}.
Within this formalism the exchange term is treated exactly within Kohn-Sham density
functional theory (KS-DFT). The resulting exchange potential is self-interaction 
free and depends explicitly on the KS orbitals, which are two component Pauli 
spinors. This formalism leads to coupled integral equations for the exchange 
potential and magnetic field. In order to avoid an expensive set of fixed spin 
moment calculations this method is needed even for the study of collinear 
systems \cite{shar05}.
In the present work we use an all electron full-potential linearized augmented
plane-wave (FP-LAPW) method for the self-consistent EXX calculations.
In this scheme, the unit cell is divided into non overlapping muffin-tin spheres 
around the atomic nuclei and a remaining interstitial region. The corresponding 
wave functions are expressed as linear combinations of atomic like functions inside 
the spheres and plane waves in the interstitial region.
The calculations used a $RK_{max} = 7$ cutoff criterion for the number of plane 
waves in the basis. The wave function is expanded up to $l=10$ and the potentials 
up to $l=6$. All calculations are done for the experimental lattice parameter of 
5.496 au. We have used a mesh of 64 {\bf k}-points in the irreducible part of the 
Brillouin zone.

\section{Results}

\begin{table}
\caption{\label{tab:mom} Magnetic moment of FeAl in Bohr magneton. The FP-EXX and 
FP-LSDA are our results. The LDA+U results from Refs. \onlinecite{mohn01}  and 
\onlinecite{pet03} are for U=5eV. The DMFT results are from Ref. \onlinecite{pet03} 
and use U=2eV.}
\begin{ruledtabular}
\begin{tabular}{cccccc}
Expt. & FP-EXX & FP-LSDA & LDA+U\cite{mohn01} & LDA+U\cite{pet03} & DMFT \\ \hline
  ?   & 0.70   & 0.69   &  0                 & 0.70               & 0    \\
\end{tabular}
\end{ruledtabular}
\end{table}

The EXX calculations find a magnetic ground state with a moment of
0.70 $\mu_B$, very close to the LSDA value 0.69$\mu_B$. It would be interesting at 
this point to compare the density of states (DOS) for FeAl with the experiments. 
A comparison between the magnetic DOS and the experimental negative binding energy 
results of Shabanova {\it et. al} will not be meaningful, 
since, as has been discussed above, the Fe sublattice is non-magnetic in experiment. 
However, since only 2\%-3\% of the Fe atoms in the FeAl sample will actually be high
moment antisite defect atoms, their contribution to the binding energy signal will
be quite small. Hence a rather good approximation to model the experimental results
will be to simply make the comparison with non-magnetic FeAl. In principle, of course,
one would wish to include the complex defect structure within the EXX approach, 
however, since this requires either very large supercells or the use of the CPA, 
this is not feasible at present.

In the lower panel of Fig. 1 we show the results of an EXX
calculation of the DOS. A most remarkable feature can immediately be seen which 
is the shift downwards of the non-bonding $t_{2g}$ states by almost 1.5 eV. This 
is quite different from the results obtained using the LDA+U \cite{mohn01} and the 
DMFT \cite{pet03} methods. Like the LDA+U method we see a broadening of the $d$-band,
in contrast to the DMFT density of states, where the $d$-band is narrowed by the 
mass renormalization effects. This spread of results by different methods makes
a comparison with experiments most interesting. In the upper panel of Fig. 1
we show the LDA and EXX DOS, as well as that reported in Refs. \onlinecite{mohn01}
and \onlinecite{pet03}. Note that for ease of comparison we have convolved our 
DOS with a Gaussian of width 0.5 eV. Clearly, the peak of the experimental DOS 
and that given by EXX are in much better agreement than any other method. A further 
interesting feature of the EXX DOS is a pronounced anti-bonding $e_g$ peak at 
around 4 eV, a feature that should be detectable in inverse photoemission experiments.

This enhanced splitting of the t$_{2g}$ and e$_g$ manifolds over LSDA values is 
the result of the asymmetry in the EXX potential. 
This is demonstrated in Fig. 2 where the EXX exchange 
potential (black) along with the LDA exchange potential (red) are displayed. A qualitative
difference between the two can be seen in that the EXX potential is strongly
asymmetric around the Fe atoms. Looking at the potential along an
Al-Fe-Fe path, indicated by the arrows in Fig. 2, 
one notes structures reflecting the existence of atomic shells. These features 
are a consequence of the natural orbital dependence of the EXX potential, but 
are completely absent in the LDA potential which is constructed from only local 
knowledge of the density. Dips in the EXX potential will mean an effective 
attractive force which will act to cause localization, the pronounced dip in the
Fe-Fe direction, for example, is most likely responsible for the sharp
$e_g$ peak in the EXX DOS.

\begin{figure}[floatfix]
\begin{center}
\includegraphics[angle=-90,width=0.50\textwidth]{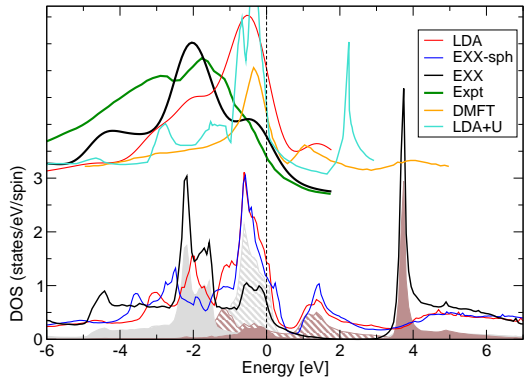}
\end{center}
\caption{Total density of states for FeAl. The LDA results are shown in red, EXX 
results with only $l=0$ part of potential in blue, and full potential EXX in black. 
The t$_{2g}$ (gray) and e$_g$ (brown) manifolds are shown with solid shading for 
EXX and hatched shading for LDA. At the top the green curve is the experimental
negative binding energy from Ref. \cite{shab01} alongside the theoretical
Gaussian convolved DOS. The cyan curve is the data extracted from Ref. 
\onlinecite{mohn01} and orange from Ref. \onlinecite{pet03}.}
\end{figure}

The importance of this asymmetry in the EXX potential may be gauged by
retaining only the spherical part in the muffin-tin spheres during
a self-consistent calculation. The resulting
EXX potential (blue line, Fig. 2 then lacks the
asymmetry of the full EXX potential, and the resulting DOS is much
closer to that of the LDA, and in disagreement with experiment.

This agreement of the EXX DOS with experiment can be considered 
somewhat surprising in that previous EXX calculations have not found this to
be so for the transition metals Fe, Co, and Ni. Here a quite striking disagreement
with experiment, as well as the LSDA, was noted \cite{kot97}. 
Although these calculations used the atomic sphere approximation,
as well as an inaccurate formulation of EXX for magnetism, our results are 
qualitatively similar\cite{shar05,shar05a}.
One finds the splitting of the $t_{2g}$ and $e_g$ states for the partially occupied 
spin down channel is far to large, with the occupied spin up channel being pulled up to
3 eV below the Fermi level.
One can speculate upon the reason for the good agreement between EXX and experimental DOS
found in this work. Since one of the main effects of the missing correlation will be simply
to screen the exchange field, one can note that the presence of Al atoms in this
material leads to an effective screening by the delocalized $sp$ Al states. The
validity of this proposal will be investigated in a future work on screened EXX. 

\begin{figure}[floatfix]
\caption{Upper image shows contour plots of the EXX (black) and LDA (red)
exchange potential in the [110] plane over two unit cells with contour
spacing 3.6 eV. In the lower graph, the same potentials are plotted along
the line indicated in the upper graph, with the blue curve corresponding
to the $l=0$ EXX. An offset is added to the EXX potentials to bring them
into coincidence with the LDA potential.
}
%\vspace{7mm}
\begin{center}
\includegraphics[angle=-00,width=0.50\textwidth]{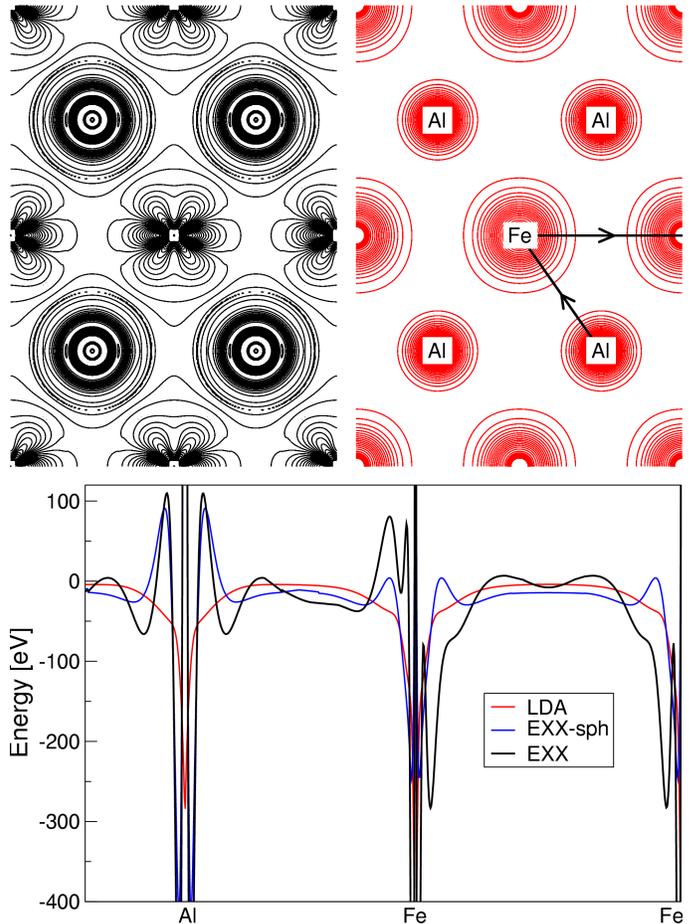}
\end{center}
\end{figure}

In conclusion, we find that FeAl prefers to be ferromagnetic in both EXX and 
LSDA calculations. However, the non-magnetic EXX DOS is in very good agreement with
experiment, while the non-magnetic LSDA DOS is not. Thus are calculations are
in agreement with the suggestion of Smirnov {\it et al.} that the non-magnetic
state of the Fe sublattice results from the complex defect structure.
However, it is also clear that the LSDA exchange and correlation is not sufficient to 
reproduce (or understand) all the properties of this interesting system.

\end{document}